\documentclass[12pt]{article}
\usepackage{epsf}
\usepackage{amsmath}
\usepackage{graphics}
\usepackage{graphicx}
\usepackage{cite}

\renewcommand{\thefootnote}{\#\arabic{footnote}}

\newcommand{\lesssim}{ \mathop{}_{\textstyle \sim}^{\textstyle <} }

\begin{document}

\setcounter{footnote}{0}
\begin{titlepage}

\begin{center}

\hfill astro-ph/0612739\\
\hfill December 2006\\

\vskip .5in

{\Large \bf
Dark Energy Parametrizations \\
and the Curvature of the Universe
}

\vskip .45in

{\large
Kazuhide Ichikawa$^{1}$ and Tomo Takahashi$^{2}$
}

\vskip .45in

{\em
$^{1}$Institute for Cosmic Ray Research,
University of Tokyo, \\
Kashiwa 277-8582, Japan \\
$^{2}$Department of physics, Saga University, Saga 840-8502, Japan
}

\end{center}

\vskip .4in

\begin{abstract}

  We investigate observational constraints on the curvature of the universe 
  not restricting ourselves to a cosmological constant as dark energy,
  in particular allowing a dark energy equation of state to
  evolve with time in several ways.
  We use type Ia supernovae (SNeIa) data from the
  latest gold data set which includes 182 SNeIa, along with the cosmic microwave background shift parameter 
  and the baryon acoustic oscillation peak. 
  We show quantitatively that the constraint on the curvature of
  the universe depends on dark energy model: 
  some popular parametrizations give constraints closely around the flat universe
  at 5\% level (2$\sigma$ C.L.) whereas some parametrizations  allow the universe
  to be as open as $\Omega_k \sim 0.2$.

\end{abstract}

\end{titlepage}

\renewcommand{\thepage}{\arabic{page}}
\setcounter{page}{1}
\renewcommand{\thefootnote}{\#\arabic{footnote}}
\renewcommand{\theequation}{\thesection.\arabic{equation}}

\section{Introduction}\label{sec:intro}
\setcounter{equation}{0}

There is now a large body of evidence indicating that the universe is
accelerating today. Dark energy is often assumed to explain the
present cosmic acceleration and has been extensively studied.  Although a lot
of models for dark energy have been proposed and investigated, we
still do not know the nature of dark energy. 
Thus phenomenological investigations have also been widely conducted by
parametrizing the dark energy equation of state $w_X$. 
Constraints on $w_X$ have been obtained from cosmological observations such as
those of the cosmic microwave background (CMB), type Ia supernovae (SNeIa), large
scale structure and so on. When one tries to constrain $w_X$, 
the simplest case would be to assume constancy in time, but most of
the recent analyses accommodate time variation of $w_X$ in some way
as predicted by many models of dark energy.

We note that a flat universe is usually assumed when constraining the time 
dependence of $w_X$. 
The assumption is often justified by invoking a prediction of the inflation
or by resorting to confirmation by cosmological observations.
However, we should test the inflationary paradigm by measuring the curvature
of the universe, and observational evidence of a flat universe is often
obtained assuming  a cosmological constant for dark energy.
Therefore, not knowing the nature of the dark energy, it is important to investigate
the curvature of the universe with various dark energy models.

In fact, when we allow the possibility of a non-flat universe, it has been
pointed out that there are some degeneracies in the CMB power spectrum 
between the curvature of
the universe and the equation of state of dark energy even if we
assume a constant equation of state \cite{Crooks:2003pa}. Hence if we
consider the possibilities of a non-flat universe and a time-varying
equation of state simultaneously, the degeneracy can become much
worse. Theoretical considerations on this curvature-dark energy degeneracy 
can be found in Refs.~\cite{Balbi:2003en,Knox:2005hx,Polarski:2005jr,Linder:2005nh,Dick:2006ev,
Huang:2006er,Knox:2006ux,Smith:2006nk,Hu:2006nm}. 
Analyses based on the recent observational data sets have been investigated in 
Refs.~\cite{Gong:2005de,Ichikawa:2005nb,Ichikawa:2006qb}.

In Ref.~\cite{Ichikawa:2005nb}, a simple time dependence of an
equation of state as $w_X= w_0 + (1-a) w_1$, with $a$ being the scale factor, was assumed and the
possibility of a
non-flat universe was considered simultaneously. The constraints
were obtained from observations of CMB, SNeIa and baryon acoustic
oscillation (BAO). It was shown that, even if we assume the
time-varying equation of state, the curvature of the universe is
constrained to be around a flat one. In fact, when a
time-varying $w_X$ is considered, the constraint on the curvature of
the universe from a single kind of observation is significantly relaxed compared to
that obtained assuming a cosmological constant. However, if we
combine three different sorts of observations, such a degeneracy
is removed, giving a tight constraint on the curvature of the universe.

In Ref.~\cite{Ichikawa:2006qb}, another parametrization was considered
to investigate the above issue. We have shown that, in a certain
parametrization, the constraint on the curvature of the universe
becomes much less stringent even if we combine the different cosmological data.
  In particular, the allowed region for the curvature of the universe
extends to the region of an open universe. In other words, it was shown that the
constraint on the curvature of the universe can depend on a model of
dark energy\footnote{
Similar analysis was done for the DGP model where it was shown that 
an open universe is slightly favored \cite{Maartens:2006yt}.
}. 

In this paper, we reinvestigate the issue of determining the curvature of the universe
with various dark energy parametrizations in the light of the recently released data set
which includes the 13 SNeIa with $z> 1$ newly discovered using the Hubble Space Telescope
\cite{Riess:2006fw}\footnote{
For recent works which use the new data, see Refs.~\cite{Barger:2006vc,Gong:2006gs,Alam:2006kj,Dutta:2006cf,Nesseris:2006ey,Zhao:2006qg}.
}. 
We show that, for some models of dark energy,  the curvature of the universe is
severely constrained to be around the flat case from three observations
combined even though the evolution of dark energy equation of state is
allowed to vary in time as found in Ref.~\cite{Ichikawa:2005nb}. 
However, if we adopt another parametrization for
$w_X$ of some type, we do not obtain a severe constraint on the
curvature of the universe but rather the region of an open universe is
largely allowed.  We also discuss what kinds of dark energy models allow
the universe to be non-flat, briefly.

The organization of this paper is as follows.  In the next section, we
summarize the parametrizations for dark energy adopted in this
paper. Then, in section \ref{sec:method}, the analysis method is briefly
explained. In section \ref{sec:constraint}, we present our result for
the constraint on the curvature of the
universe for several parametrizations of the dark energy equation of state.
In the final section, we give conclusion and discussion.

\section{Parametrizations of dark energy} \label{sec:params}

In this section, we summarize the parametrizations for dark energy
equation of state adopted in this paper.

When one tries to accommodate a time-varying equation of state, one of the
simplest parametrizations may be the one which adds a linear dependence on the scale
factor $a$ as follows \cite{Chevallier:2000qy,Linder:2002et}:
\begin{equation}
\hspace{-3cm} {\rm Parametrization\ A}: \quad w_X   = w_0 + (1-a) w_1 = w_0 + \frac{z}{1+z} w_1,
\label{eq:eos1}
\end{equation}
where $z$ is the redshift.  We call this 
parametrization A in this paper.  In this parametrization, the
equation of state becomes $w_X = w_0$ at the present time and $w_X =
w_0 + w_1$ at earlier time.  If $w_X$ becomes larger than 0, it means
that the energy density of dark energy decreases faster than that of
matter. Since we investigate the case where dark energy drives the
late time cosmic acceleration, we do not consider such a  possibility, 
which can be taken into account by assuming
\begin{equation}
w_0+w_1 < 0. 
\label{eq:w0w1}
\end{equation}
Notice that this prior of negative $w_X$ is always assumed in the following analysis.
The energy density for dark energy with the parametrization of
Eq.~\eqref{eq:eos1} can be written analytically as
\begin{equation}
\rho_X(z)  
=  \rho_{X0} 
(1+z)^{3(1+w_0+w_1)}  \exp \left( \frac{-3 w_1 z}{1+z} \right),
\end{equation}
where $\rho_{X0}$ is the energy density of dark energy at the present
time.

The next parametrization for dark energy used in this paper is the
following one:
\begin{equation}
\hspace{-3cm} {\rm Parametrization\ B}: \quad w_X(z) = 
\begin{cases}
 w_0 + \displaystyle\frac{w_1 - w_0}{z_*} z 
 &( {\rm for}~~ z \le z_*)  \\ \\
 w_1      & ({\rm for}~~ z \ge z_*),   
\end{cases} 
\label{eq:eos2}
\end{equation}
where we interpolate $w_X$ linearly from the present epoch to a
redshift $z_*$ to which we refer as the transition redshift.
We call this parametrization B.  The
value of the equation of state becomes $w_X(z=0) = w_0$ today and
$w_X( z \ge z_*) = w_1$ before the transition redshift.  The energy
density of dark energy can be written as
\begin{equation}
\rho_X (z) =  \rho_{X0} \times
\begin{cases}
 \exp [3 \alpha z] (1+z)^{3(1 + w_0 - \alpha)} 
&(  {\rm for}~~ z \le z_*)  \\ \\
 \exp [3 \alpha z_*] (1+z_*)^{3(1 + w_0 - \alpha)} 
\displaystyle\left( 
\frac{1+z}{1+z_*}\right)^{3(1+w_1)}
& ({\rm for}~~ z \ge z_*),   
\end{cases} 
\end{equation}
where 
\begin{eqnarray}
\alpha & \equiv & \frac{w_1 - w_0}{z_*}.
\end{eqnarray}
In fact, this parametrization is essentially the same as the one with
$w_X(z) = w_0 + \alpha z$ with a cut-off at some redshift to avoid a
large value of $w_X$ at early times, which has been adopted in the
literatures {\it e.g.} Refs.~\cite{Huterer:2000mj,Weller:2001gf,Frampton:2002vv}. In
the following analysis, we consider several values of $z_*$ fixed, then
vary other parameters.

The third parametrization for a dark energy equation of state adopted
in the following is the one proposed in Ref.~\cite{Wetterich:2004pv}
which is written as
\begin{equation}
\hspace{-5.5cm} {\rm Parametrization\ C}: \quad w_X (z) = \frac{w_0}{[1 + b \log (1+z) ]^2},
\label{eq:eos3}
\end{equation}
where $w_0$ represents the value of $w_X$ at the present time which
should be negative in order to realize the current cosmic
acceleration.  The value of $b$ is assumed to be positive to avoid a
singularity of $w_X \rightarrow -\infty$ at some redshift for $1 + z
\ge 0$. Thus, as the redshift increases, the value of $w_X$
approaches $w_X \rightarrow 0$ which can include the possibilities
that dark energy contributes to the total energy of the universe to
some extent at an earlier epoch. In fact, this parametrization is
motivated to include such an early time dark energy
\cite{Wetterich:2004pv}.  We call this parametrization C.
The energy density of dark energy
for this parametrization can be written as
\begin{equation}
\rho_X (z) = \rho_{X0}    (1 + z )^{3 + 3 \tilde{w}_X(z)},
\label{eq:wetterich_rho}
\end{equation}
where $\tilde{w}_X(z) = w_0 /[ 1 + b \log (1+z) ]$. 

After specifying the model and its parameters, we can calculate the
evolution of the Hubble parameter using
\begin{equation}
H^2(z) = H_0^2 \left[ 
\Omega_r (1+z)^4 + \Omega_m(1+z)^3  + \Omega_k(1+z)^2 
+ \Omega_X \exp \left( 
3 \int_0^z ( 1 + w_X(\bar{z})) \frac{d\bar{z}}{1+\bar{z}} \right) 
\right],
\end{equation}
where the integration in the dark energy density can be analytically performed in our cases 
as mentioned above. This in turn is used to compute cosmological distances 
which are introduced in Sec.~\ref{sec:method}.

\section{Analysis method} \label{sec:method}

In this section, we briefly discuss the analysis method and cosmological data used in this paper.

For the data from SNeIa, we fit the distance modulus calculated in a
dark energy model to the observational data released recently \cite{Riess:2006fw}. 
We use 182 SNeIa from the gold data provided in \cite{R06data}.
The distance modulus can be calculated as
\begin{equation}
M -m = 5\log d_L + 25. 
\end{equation}
Here $d_L$ is the luminosity distance in units of Mpc which is written as
\begin{equation}
d_L = \frac{1+z}{H_0 \sqrt{|\Omega_k|} }
\mathcal{S}  \left( \sqrt{|\Omega_k|} \int_0^z \frac{dz'}{H(z') /H_0} \right),
\end{equation}
where $\mathcal{S}$ is defined as $\mathcal{S}(x) = \sin (x)$ for a
closed universe, $\mathcal{S}(x) = \sinh (x)$ for an open universe and
$\mathcal{S}(x) = x$ with the factor $\sqrt{|\Omega_k|}$ being removed
for a flat universe.

For the CMB data, we make use of the shift parameter which is a good
measure of the position of the first peak of the CMB power spectrum which can be
written as
\begin{equation}
R  = \frac{\sqrt{\Omega_m}}{\sqrt{|\Omega_k|} }
\mathcal{S}  \left( \sqrt{|\Omega_k|} \int_0^{z_{\rm rec}} \frac{dz'}{H(z')/H_0} \right),
\end{equation}
where $z_{\rm rec}=1089$ is the redshift of the epoch of the recombination.
From the three-year WMAP result
\cite{Spergel:2006hy,Page:2006hz,Hinshaw:2006ia,Jarosik:2006ib}, the
shift parameter is constrained to be $R = 1.70 \pm 0.03$
\cite{Wang:2006ts}.  Since here we consider the shift parameter which
is determined only by the background evolution for the constraint from
CMB, we do not need to include the effect of the fluctuation of dark
energy. In this paper, by using a shift parameter, we can confine
ourselves to considering the effects of the modification of the
background evolution alone.

We also include the data of BAO making use of the so-called 
parameter $A$:
\begin{equation}
A = \frac{\sqrt{\Omega_m}}{(H(z_1)/H_0)^{1/3}} \left[  
\frac{1}{z_1 \sqrt{|\Omega_k|} }
\mathcal{S}  \left( \sqrt{|\Omega_k|} \int_0^{z_1} \frac{dz'}{H(z')/H_0} \right)
\right]^{2/3},
\end{equation} 
where $z_1=0.35$ and $A$ is measured as $A=0.469 (n_s/0.98)^{-0.35}
\pm 0.017$ \cite{Eisenstein:2005su}.  Here the dependence on the
scalar spectral index is shown explicitly. For the analysis in this
paper, we adopt $n_s=0.95$ which is the mean value for $\Lambda$CDM
model from WMAP3 data alone  \cite{Spergel:2006hy}. (Since the main effect of dark energy equation of 
state on the CMB power spectrum is just shifting the acoustic peak positions,
$n_s$ would be similar value even if we did not assume a cosmological constant.
Actually, the mean value for the model with a constant equation of state of dark energy 
is $n_s =0.954$ \cite{Spergel:2006hy}. Thus the use of $\Lambda$CDM mean value
should not affect our quantitative results.) 
  
We present constraints from all three observations combined in the next section.

\section{Constraints from recent observations} \label{sec:constraint}

\begin{figure}[htb]
\begin{center}
\vspace{-12cm}
\scalebox{0.6}{\includegraphics{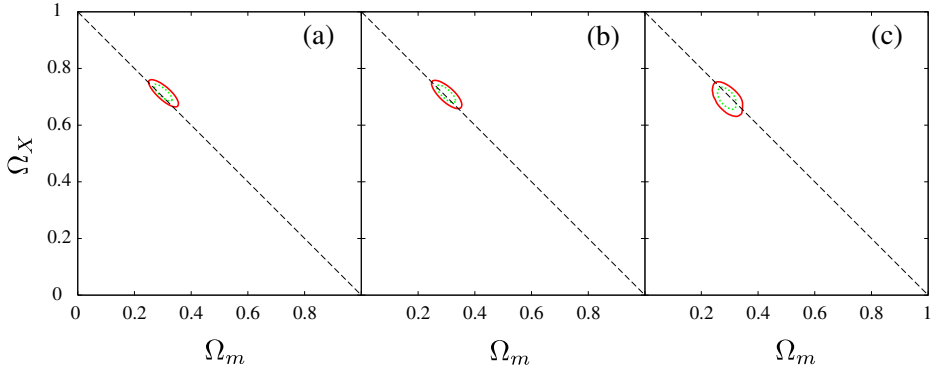}}
\end{center}
\caption{Contours of 1$\sigma$ and 2$\sigma$ allowed regions
  are shown on the $\Omega_m$-$\Omega_X$ plane for the parametrization
  A (panel (c)).  For comparison, the cases with a cosmological
  constant (a) and a constant equation of state (b) are also presented.  }
\label{fig:A}
\end{figure}

\begin{figure}[htb]
\begin{center}
\vspace{-9cm}
\scalebox{0.6}{\includegraphics{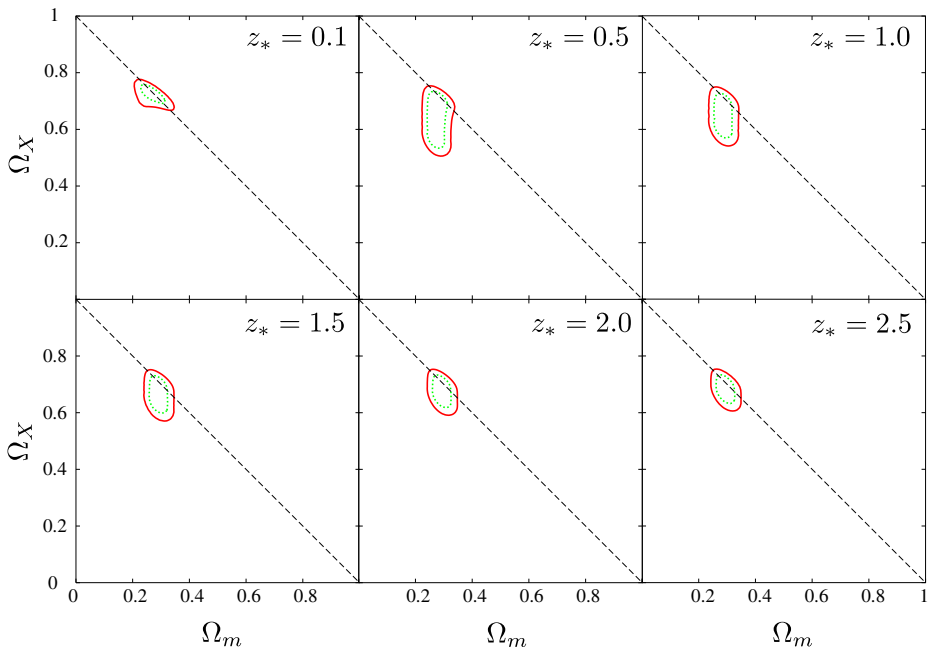}}
\end{center}
\caption{Contours of 1$\sigma$ and 2$\sigma$  allowed regions are shown 
on the $\Omega_m$-$\Omega_X$ plane for the parametrization B.}
\label{fig:B}
\end{figure}

\begin{figure}[htb]
\begin{center}
\vspace{-12cm}
\scalebox{0.6}{\hspace{5.5cm} \includegraphics{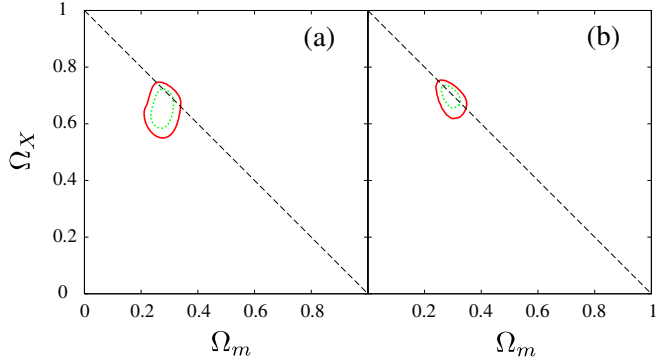}}
\end{center}
\caption{Contours of 1$\sigma$ and 2$\sigma$ allowed regions
  are shown on the $\Omega_m$-$\Omega_X$ plane for the parametrization
  C.  Two cases with different priors on $w_0$ are shown: (a) $-5 \le
  w_0 \le -0.5$ and (b) $-1 \le w_0 \le -0.5$.}
\label{fig:C}
\end{figure}

In this section, we show constraints on the curvature of the universe for several 
parametrizations of the time evolution of dark energy equation of state introduced
 in Sec.~\ref{sec:params}.
We present the results in Figs.~\ref{fig:A}-\ref{fig:C}, drawing
1$\sigma$ and 2$\sigma$ C.L. contours in the $\Omega_m$-$\Omega_X$ plane 
from the combination of all three observational data sets which are explained
 in Sec.~\ref{sec:method}. 2$\sigma$ bounds on $\Omega_k$ and best fit
 parameters are summarized in Table \ref{tab:bestparam}.

Fig.~\ref{fig:A} shows the constraint for the parametrization A (Eq.~\eqref{eq:eos1}) as well as
the cases with a cosmological constant and a constant equation of state
(in terms of the parametrization A, a cosmological constant can be regarded as 
the case with $w_0=-1$ and $w_1=0$ and a constant equation of state as for  the $w_1=0$ case).  
For the parametrization A, we marginalize over $w_0$ and $w_1$ and for 
a constant equation of state, $w_0$ is marginalized over. As 
seen from Fig.~\ref{fig:A}, the allowed regions for three cases lie
closely around a flat universe (denoted by the straight line $\Omega_m+\Omega_X=1$).
This means that, even if we drop the assumption of a cosmological constant
for dark energy, 
as long as the dark energy equation of state is constant or
its time variation is expressed as the parametrization A, 
the curvature of the universe is well constrained
to be around the flat case (although the allowed region for the case with the
parametrization A is slightly larger than those for the other
cases).  In fact, this finding was already made in
Ref.~\cite{Ichikawa:2005nb}, which is reconfirmed by the analysis
here with updated data sets. 
Notice that, of course, a single kind of observation has severe degeneracy
among the curvature of the universe and the dark
energy parameters $w_0$ and $w_1$.  However, such a degeneracy is removed
when we use three data combined as discussed in
Ref.~\cite{Ichikawa:2005nb}. 
We also compute other constraints such
as the one in the $\Omega_m$-$w_0$ plane, the $w_0$-$w_1$ plane and so on using
updated data and find that they are almost unchanged from those of the 
previous analysis of Ref.~\cite{Ichikawa:2005nb}.

Next we show the result for the parametrization B (Eq.~\eqref{eq:eos2}) in Fig.~\ref{fig:B}. 
It has three model parameters in the equation of
state, $w_0$, $w_1$ and $z_*$. Here we report our analysis, fixing the
value of $z_*$ to several values to see how the transition redshift
affects the constraint on the curvature of the universe, which is, as
it turns out, important for understanding what type of dark energy allows a
considerably non-flat universe. In Fig.~\ref{fig:B}, the constraints for 
$z_*= 0.1, 0.5, 1.0, 1.5, 2.0$ and $2.5$ are shown. We marginalize
over $w_0$ and $w_1$.
The constraint on the curvature of the universe
for this parametrization was already discussed in
Ref.~\cite{Ichikawa:2006qb}
where it was shown that a significantly large region of
an open universe is allowed for some particular transition redshift
$z_*$.  Although the analysis in the present
paper includes more SNeIa with high redshift $z > 1$ than were used in
Ref.~\cite{Ichikawa:2005nb}, the general results on the constraint on
the curvature are almost unchanged.  We can see that
the curvature is constrained to be around
the flat case to some extent when the transition redshift is as small as $z_* \sim
0.1$ or as large as $z_* > 2.0$.  This is because, when the transition
redshift is small, such a model becomes like the one with a constant
$w_X$. In this case, the curvature is well constrained to be around
the flat case as similar to the case in panel (b) of Fig.~\ref{fig:A}.  On the other
hand, when $z_*$ is large, this model behaves like the parametrization A; 
thus the constraint is similar to that of the case in
panel (c) of Fig.~\ref{fig:A}, where the curvature is also
constrained to be around the flat case.
However, when the transition redshift
is in an intermediate range such as $ 0.5 \lesssim z_* \lesssim 1.5$,
the allowed region includes a larger region of an open universe. The
parameters which give the best fit to the data for the case with $z_*
= 0.5$ are $w_0 = -1.40$ and $w_1 = -0.32$.
 Typically a model which gives 
a good fit to the data has its
equation of state being smaller than $-1$ at the present and nearly 0
in the past. Thus a model of dark energy
which allows an open universe is like one which behaves
similarly to the matter at earlier time, but at present time, its
equation of state becomes very small.  Furthermore, it should be
mentioned that the minimum value of the total $\chi^2$ for the case
with $z_*= 0.5$ becomes smaller than that of the $\Lambda$CDM case by
4.1, which means that this parametrization can give a better fit to
the data than the $\Lambda$CDM model.  
The current cosmological data cannot give a stringent constraint on 
the curvature of the universe when we assume this kind of dark energy model,
as was concluded in Ref.~\cite{Ichikawa:2005nb}.

Now we are going to discuss the case with the parametrization C (Eq.~\eqref{eq:eos3}).
 We show the constraint for this model in Fig.~\ref{fig:C} with two different priors. 
The first prior is somewhat generic: $-5 \le w_0 \le -0.5$ (panel (a) in
Fig.~\ref{fig:C}). The other prior is meant to avoid a
so-called phantom dark energy: $-1 \le w_0 \le -0.5$ (panel (b) in
Fig.~\ref{fig:C}). We marginalize over $w_0$ in these ranges and also over $b$
accordingly. When we adopt the first prior, the allowed
parameter range in the $\Omega_m$-$\Omega_X$ plane occupies a relatively large
region of an open universe as broad as for the case of the parametrization B
with the intermediate transition redshift. The $\chi^2_{\rm min}$ can be lowered 
by 3.9 compared to the cosmological constant case.
On the other hand, when we forbid phantom dark energy by
adopting the second prior, the constraint on the curvature of the universe becomes more severe.
As mentioned above for the result of the parameterization B,
the dark energy model which allows a large region of an open universe
has $w_X <-1$ at present epoch and $w_X \sim 0$ with an appropriate transition redshift.
It is easy to see the parametrization C can realize these conditions since it has $w_X \sim 0$
in the past by definition and we can tune the present value by varying $w_0$ and the transition redshift 
to some extent by varying $b$. If $w_0$ is not permitted to go below $-1$, the fit for
an open universe cannot be as good as that for the phantom case, so the allowed region is limited
in the neighborhood of a flat universe.

\begin{table}[htb]
  \centering 
  \begin{tabular}{|c||c|c|c|c|c|c|}
  \hline
  & $\Omega_k$ ($2\sigma$ limit) & $\chi^2_{\rm min}$ & $\Omega_m$ & $\Omega_X$ & $w_0$ & $w_1$ \\
  \hline
  Cosmological constant  & $[-0.053,0.018]$ &160.8 & 0.30 & 0.71 & $-$ & $-$ \\
  Constant $w_X$ & $[-0.047,0.027]$ &159.6 & 0.30 & 0.71 & -0.90 & $-$\\
   Parametrization A  & $[-0.042,0.074]$ &157.9 & 0.29 & 0.69 & -1.12 & 1.12 \\
   \hline
  \hline
Parametrization   B &  & &  &  & $w_0$ & $w_1$ \\
\hline
$z_\ast = 0.1$ & $[-0.044,0.080]$ &157.9 & 0.26 & 0.73 & -2.75 & -0.64 \\
  $z_\ast =0.5 $ & $[-0.036, 0.224]$ &156.7 &0.27 & 0.69 & -1.40 & -0.32 \\
 $z_\ast =1.0 $ & $[-0.038, 0.175]$ &157.2 & 0.28 & 0.64 & -1.25 & -0.06 \\
 $z_\ast =1.5 $ & $[-0.044, 0.139]$ &157.9 & 0.29 & 0.64 & -1.15 & ~0.00 \\
 $z_\ast =2.0 $ & $[-0.044,0.114] $&158.3 & 0.29 & 0.66 & -1.06 & -0.04\\
 $z_\ast =2.5 $ & $[-0.044,0.098]$ &158.6 & 0.30 & 0.67 & -1.04 & ~0.00\\
   \hline
   \hline
 Parametrization  C &    &    &  & & $w_0$ & $b$ \\
\hline
& $[-0.036,0.193]$ & 156.9 & 0.27 & 0.64 & -1.94 & ~2.70 \\
 $w_X \ge -1$ & $[-0.043,0.092]$ &158.6 & 0.30 & 0.70 & -1.00 & ~0.30 \\
   \hline
  \end{tabular}
  \caption{Comparison of the 2$\sigma$ constraints on the curvature of the universe $\Omega_k$ for 
  the parametrizations adopted in this paper (see Sec.~\ref{sec:params} for their definition). We also show the minimum $\chi^2$ values and best-fit parameters.}
  \label{tab:bestparam}
\end{table}

\begin{figure}
\begin{center}
\vspace{-15cm}
\scalebox{0.8}{\hspace{3cm}\includegraphics{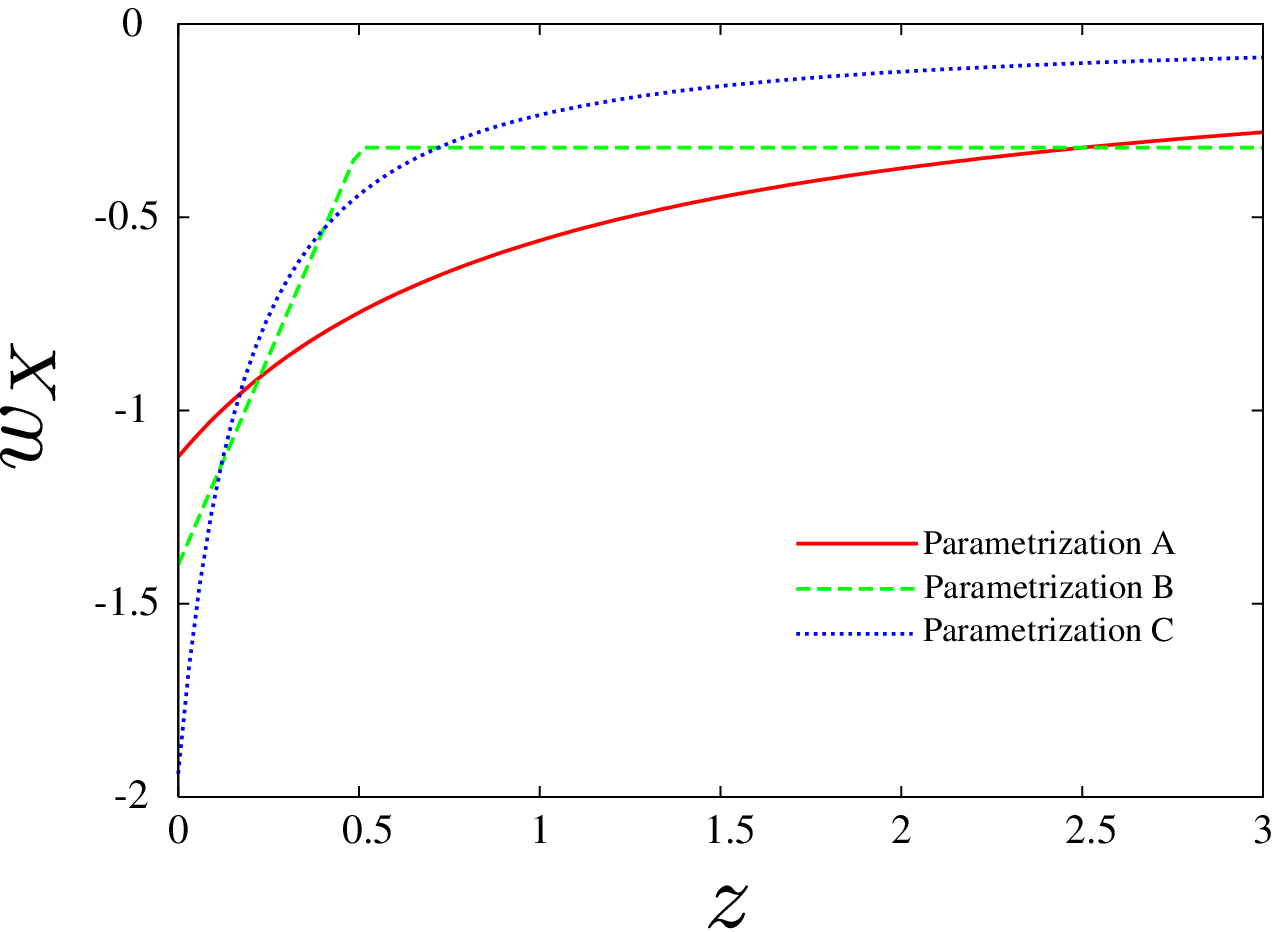}}
\end{center}
\caption{Evolutions of $w_X$ for the parametrizations A (red solid line), B (green dashed line) and C (blue dotted line). We choose parameters which give minimum $\chi^2$ for each parametrization as reported in Table~\ref{tab:bestparam}.}
\label{fig:w_evolution}
\end{figure}

The similarity of the parametrizations B and C may be visually understood as follows. Fig.~\ref{fig:w_evolution} shows how $w_X$ varies with respect to $z$ for the parametrizations A, B and C. We choose parameters which give minimum $\chi^2$ for each parametrization, namely, $w_0=-1.12$ and $w_1 = 1.12$ for the parametrization A, $z_{\ast} = 0.5$, $w_0=-1.40$ and $w_1=-0.32$ for the parametrization B, and $w_0=-1.94$ and $b=2.70$ for the parametrization C, as reported in Table~\ref{tab:bestparam}. The resemblance of cases B and C is apparent, especially as regards the transition redshift around $z\sim 0.5$. By contrast, parametrization A cannot give rise to such a transition as its functional form only allows a mild transition around $z\sim 1$. 

\section{Conclusion and discussion} \label{sec:conclusion}

We discussed the issue of the determining the curvature of the
universe, considering several types of dark energy model, in particular ones 
with a time-varying equation of state. 
Usually a
flat universe is assumed when we give constraints on dark energy
models. This assumption looks reasonable since it is often said that
current cosmological data favor a flat universe. However, notice that
when one does the analysis to constrain the curvature of the universe,
a cosmological constant is assumed for dark energy in most cases. Thus
it is not so obvious how the constraint on the curvature changes when
we remove the assumption of a cosmological constant for dark energy.
In this paper, the constraints on the curvature are investigated
without this assumption, using  the recent SNeIa data of the gold data set
\cite{Riess:2006fw}, the CMB shift parameter from WMAP3 and BAO.

We have analyzed three kinds of dark energy parametrization which are
denoted as the parametrizations A, B and C here (Eqs.~\eqref{eq:eos1}, \eqref{eq:eos2} and \eqref{eq:eos3}
respectively).  The first parametrization
A has been used in many literature since this is a simple way to
include the time evolution of a dark energy equation of state.  We
have shown that the curvature of the universe is well constrained to
be around the flat case when we assume the parametrization A, which has been already found in
Ref.~\cite{Ichikawa:2005nb}.  In this paper, we reanalyzed this model
using the recent gold data which include more SNeIa with high redshift $z
\ge 1$ and confirmed the results.

Then we have considered the parametrization B, which was also
investigated in Ref.~\cite{Ichikawa:2006qb} where earlier data were used.
In this paper, we have found that even if we include the new data from
SNeIa, the general conclusion remains the same as the one drawn in
Ref.~\cite{Ichikawa:2006qb}: for a dark energy model like the
parametrization B with a particular transition redshift, the
curvature of the universe is not so severely constrained to be around
the flat case, but rather the allowed region includes a relatively large area
of an open universe. In particular, we have discussed that 
some characteristic properties of dark energy are required for such a case. 
That is, an open universe tends to be allowed in a case where the equation of
state for a dark energy is smaller than $-1$ at the present time
and approaches to zero at earlier time with a somewhat abrupt transition at 
a redshift in the range $ 0.5\lesssim z_* \lesssim 1.5$.

The third model we considered in this paper was the one proposed in
Ref.~\cite{Wetterich:2004pv}.  One of the motivations for this
parametrization is that it can accommodate the case where dark energy
can contribute to the total energy density of the universe even at
earlier time.  Thus we can expect that this model to have the same
characteristics as the case with the parametrization B.  We showed that,
in this parametrization, the allowed region of the curvature of the
universe, from current observations, also extends to the region of an
open universe after marginalizing over the equation of state
parameters in the model.

We have explicitly shown that, for some models of dark energy, the
curvature of the universe is allowed to be open. This may have many
implications for the issue of the determination of cosmological  
parameters from observations.  Since we do not understand the nature
of dark energy yet, we cannot make any particular choice for models of
dark energy with definite criteria. Having this situation in mind, we
should be careful when we constrain the curvature of the universe. 
Furthermore, the assumption of a flat
universe should be adopted with caution for some dark energy models
if one would like to do the analysis from a general point of view.

\section*{Acknowledgment} 
We gladly acknowledge an informative correspondence with Danny Marfatia.



\begin{thebibliography}{100}
\bibitem{Crooks:2003pa}
  J.~L.~Crooks, J.~O.~Dunn, P.~H.~Frampton, H.~R.~Norton and T.~Takahashi,
  Astropart.\ Phys.\  {\bf 20}, 361 (2003)
  [arXiv:astro-ph/0305495].

\bibitem{Balbi:2003en}
  A.~Balbi, C.~Baccigalupi, F.~Perrotta, S.~Matarrese and N.~Vittorio,
  Astrophys.\ J.\  {\bf 588}, L5 (2003)
  [arXiv:astro-ph/0301192].

\bibitem{Knox:2005hx}
  L.~Knox,
  Phys.\ Rev.\ D {\bf 73}, 023503 (2006)
  [arXiv:astro-ph/0503405].

\bibitem{Polarski:2005jr}
  D.~Polarski and A.~Ranquet,
  Phys.\ Lett.\ B {\bf 627}, 1 (2005)
  [arXiv:astro-ph/0507290].

\bibitem{Linder:2005nh}
  E.~V.~Linder,
  Astropart.\ Phys.\  {\bf 24}, 391 (2005)
  [arXiv:astro-ph/0508333].

\bibitem{Dick:2006ev}
  J.~Dick, L.~Knox and M.~Chu,
  JCAP {\bf 0607}, 001 (2006)
  [arXiv:astro-ph/0603247].

\bibitem{Huang:2006er}
  Z.~Y.~Huang, B.~Wang and R.~K.~Su,
  arXiv:astro-ph/0605392.

\bibitem{Knox:2006ux}
  L.~Knox, Y.~S.~Song and H.~Zhan,
  arXiv:astro-ph/0605536.

\bibitem{Smith:2006nk}
  K.~M.~Smith, W.~Hu and M.~Kaplinghat,
  Phys.\ Rev.\ D {\bf 74}, 123002 (2006)
  [arXiv:astro-ph/0607315].

\bibitem{Hu:2006nm}
  W.~Hu, D.~Huterer and K.~M.~Smith,
  arXiv:astro-ph/0607316.

\bibitem{Gong:2005de}
  Y.~g.~Gong and Y.~Z.~Zhang,
  Phys.\ Rev.\ D {\bf 72}, 043518 (2005)
  [arXiv:astro-ph/0502262].

\bibitem{Ichikawa:2005nb}
  K.~Ichikawa and T.~Takahashi,
  Phys.\ Rev.\ D {\bf 73}, 083526 (2006)
  [arXiv:astro-ph/0511821].

\bibitem{Ichikawa:2006qb}
  K.~Ichikawa, M.~Kawasaki, T.~Sekiguchi and T.~Takahashi,
  JCAP {\bf 0612}, 005 (2006)
  [arXiv:astro-ph/0605481].

\bibitem{Zhao:2006qg}
  G.~B.~Zhao, J.~Q.~Xia, H.~Li, C.~Tao, J.~M.~Virey, Z.~H.~Zhu and X.~Zhang,
  arXiv:astro-ph/0612728.

\bibitem{Maartens:2006yt}
  R.~Maartens and E.~Majerotto,
  Phys.\ Rev.\ D {\bf 74}, 023004 (2006)
  [arXiv:astro-ph/0603353].

\bibitem{Riess:2006fw}
  A.~G.~Riess {\it et al.},
  arXiv:astro-ph/0611572.

\bibitem{Barger:2006vc}
  V.~Barger, Y.~Gao and D.~Marfatia,
  arXiv:astro-ph/0611775.
  
  \bibitem{Gong:2006gs}
  Y.~G.~Gong and A.~z.~Wang,
  arXiv:astro-ph/0612196.

\bibitem{Alam:2006kj}
  U.~Alam, V.~Sahni and A.~A.~Starobinsky,
  arXiv:astro-ph/0612381.
  
  \bibitem{Dutta:2006cf}
  K.~Dutta and L.~Sorbo,
  arXiv:astro-ph/0612457.
  
\bibitem{Nesseris:2006ey}
  S.~Nesseris and L.~Perivolaropoulos,
  arXiv:astro-ph/0612653.

\bibitem{Chevallier:2000qy}
  M.~Chevallier and D.~Polarski,
  Int.\ J.\ Mod.\ Phys.\ D {\bf 10}, 213 (2001)
  [arXiv:gr-qc/0009008].

\bibitem{Linder:2002et}
  E.~V.~Linder,
  Phys.\ Rev.\ Lett.\  {\bf 90}, 091301 (2003)
  [arXiv:astro-ph/0208512].

\bibitem{Huterer:2000mj}
  D.~Huterer and M.~S.~Turner,
  Phys.\ Rev.\ D {\bf 64}, 123527 (2001)
  [arXiv:astro-ph/0012510].

\bibitem{Weller:2001gf}
  J.~Weller and A.~Albrecht,
  Phys.\ Rev.\ D {\bf 65}, 103512 (2002)
  [arXiv:astro-ph/0106079].

\bibitem{Frampton:2002vv}
  P.~H.~Frampton and T.~Takahashi,
  Phys.\ Lett.\ B {\bf 557}, 135 (2003)
  [arXiv:astro-ph/0211544].

\bibitem{Wetterich:2004pv}
  C.~Wetterich,
  Phys.\ Lett.\ B {\bf 594}, 17 (2004)
  [arXiv:astro-ph/0403289].

\bibitem{R06data}
The gold data set from recent HST discoveries is available at:\\
http://braeburn.pha.jhu.edu/ $\tilde{}$ ariess/R06 

\bibitem{Spergel:2006hy}
    D.~N.~Spergel {\it et al.},
    arXiv:astro-ph/0603449.
  
\bibitem{Page:2006hz}
    L.~Page {\it et al.},
    arXiv:astro-ph/0603450.

\bibitem{Hinshaw:2006ia}
    G.~Hinshaw {\it et al.},
    arXiv:astro-ph/0603451.

\bibitem{Jarosik:2006ib}
    N.~Jarosik {\it et al.},
    arXiv:astro-ph/0603452.

\bibitem{Wang:2006ts}
  Y.~Wang and P.~Mukherjee,
  Astrophys.\ J.\  {\bf 650}, 1 (2006)
  [arXiv:astro-ph/0604051].

\bibitem{Eisenstein:2005su}
  D.~J.~Eisenstein {\it et al.}  [SDSS Collaboration],
  Astrophys.\ J.\  {\bf 633}, 560 (2005)
  [arXiv:astro-ph/0501171].

\end{thebibliography}
\end{document}